%% file: main.tex
\documentclass{sig-alternate-10pt}

\input{preamble}
\usepackage{multicol}
\usepackage{lscape}

\usepackage{verbatim} 
\usepackage[hyphens]{url}
\usepackage{eurosym}
\usepackage{xcolor}
\usepackage{algorithm}
\usepackage{algorithmicx}
\usepackage{algpseudocode}
\usepackage{epsfig,endnotes}
\usepackage{footnote}
\usepackage{comment}

\usepackage{dirtree,array}

\usepackage{pgfplots}
\usepackage{tikz}\usetikzlibrary{cd} 

\newcommand{\final}{\relax}
\ifdefined\final
  \newcommand{\authorsnote}[2]{}
\else
  \newcommand*{\authorsnote}[2]{\textcolor{#1}{[#2]}}
\fi

\newcommand*{\mct}   [1]{\authorsnote{magenta}{MCT: {#1}}}
\newcommand*{\mj}   [1]{\authorsnote{purple}{MJ: {#1}}}

\usepackage{array}
\usepackage[tight,footnotesize]{subfigure}
\DeclareMathSizes{10}{10}{7}{8}
\usepackage{balance}
\usepackage{tablefootnote}
\floatname{algorithm}{Procedure}

\algnewcommand{\LineComment}[1]{\State \(\triangleright\) #1}

\newcommand{\unchecked}{\colorbox{yellow!30}}
\newcommand{\unverifiedblock}{\colorbox{orange!30}}                             
\newcommand{\crawlerblock}{\colorbox{green!30}}
\newcommand{\trueblock}{\colorbox{red!40}}

\begin{document}

\title{Exploring Server-side Blocking of Regions}

\author{Sadia Afroz\\ ICSI and UC Berkeley 
\and Michael Carl Tschantz\\ ICSI 
\and Shaarif Sajid\\ LUMS
\and Shoaib Asif Qazi\\ LUMS
\and Mobin Javed\\ LUMS
\and Vern Paxson\\ ICSI and UC Berkeley}

\maketitle
\begin{abstract}
One of the Internet's greatest strengths is the degree to which it
facilitates access to any of its resources from users anywhere in the world.
However, users in the developing world have complained of websites blocking
their countries.
We explore this phenomenon using a measurement study.
With a combination of automated page loads, manual checking, and traceroutes,
we can say, with high confidence, that some websites do block users from some
regions.
We cannot say, with high confidence, why, or even based on what criteria,
they do so except for in some cases where the website states a reason.
We do report qualitative evidence that fears of abuse and the costs of serving
requests to some regions may play a role.
\end{abstract}

\section{Introduction}\label{sec:introduction}

During a stay in Ghana, one of our colleagues 
noticed that she 
could not access \url{westelm.com}~\cite{burrell2012invisible}. %
 Visits to a variety of other websites revealed more unavailability: travel 
websites like Orbitz and Expedia did not permit Accra, Ghana, to be selected as 
a travel destination; her attempt to use PayPal launched a ``random'' security 
screening that required her to retrieve a code from her phone in the United States 
(inaccessible to her in Ghana) before proceeding. The dating site 
\url{plentyoffish.com} disallowed users from selecting Ghana as their country of 
residence, then later stated that they blocked all traffic from Africa~\cite{burrell2012invisible}.
Prior to 2015,
Amazon did not allow shipping to Nigeria~\cite{amazon-policy}.

In this work, we quantitatively demonstrate that such
unavailability is a measurable phenomenon.
Despite the anecdotal and qualitative evidence for it, we know of no
systematic studies of the differences in server-side blocking rates
across regions.
Rather, we find an abundance of measurements of censorship, some of
which risk counting server-side blocking motivated by reasons other
than censorship.
This risk highlights the first importance of measuring regional
differences more generally: a baseline to which to compare when
looking at more specific forms of it and for interpreting regional
differences in the unavailability of websites.

The second importance is that, as highlighted above, such regional
differences can adversely affect would-be website visitors.
This adversity can be understood both in terms of individuals and in
terms of the populations affected.
Our measurements show cases where much, if not all, of a region is
blocked.

We measure the unavailability of a website caused by that
website itself through actions or inactions of the website's server
(including CDNs working on behalf of the website's operators).
We term this \emph{server-side blocking}.
To do so, we use network measurements to find blocking at various
network layers.
We then use additional measurements to
differentiate between server-side blocking and blocking by
middleboxes (as seen with censorship).
We also characterize the methods used for server-side blocking.

Our measurements use the following steps:
\begin{enumerate}
\item Measure the availability of websites from vantages inside and
  outside of a region expected to experience regional server-side
  blocking,

\item Find region-wide patterns of availability and unavailability,

\item Manually confirm the unavailability of websites, and

\item Identify server-side blocking using traceroutes.

\end{enumerate}
The main novelty of our methods is the use of traceroute
to confirm that a server is doing the blocking whereas previously
only cases where the website volunteered this information were possible
to identify.
Along the way, we use a mix of automated and manual methods to
characterize unavailability in terms such as network errors, block
pages, \textsc{captcha}s, and more.

We ran our tests to compare to developing regions with the US.
We found 42~websites to be unavailable in Pakistan.
Of these, we were able to determine with high confidence that 14~websites 
were using server-side blocking.
Unexpectedly, a large fraction of these are websites are in India, not the developed world.
For vantages in Africa, we found 19 to be unavailable and 13 to be so from server-side blocking.

These findings are subject to false positives and false negatives.
Due to the lack of research on server-side blocking of regions, we
focus on demonstrating its existence with high confidence instead of
attempting to measure all of it.
With this in mind, we err on the side of false negatives and
underestimate the extent of such blocking, which partly explains
the small number of websites we flag as such.
For example, we manually inspect every website we claim to be engaged
in server-side blocking.
Furthermore, we focus on methods can confirm blocking as being
server-side without relying upon the website to announce this fact or
be truthful about it.
This requires time-consuming traceroutes, which also limits the number
of websites we identify.

One may wish to drawn conclusions about issues such as censorship or
discrimination from our findings.
We offer caution here, not only because these terms are vague,
contested, and nominative, but also because our network-level
measurements can at best serve as proxies for most common definitions
of these concepts.
For example, censorship typically describes a third-party interfering
with communications between willing participants and includes
server-side blocking due to government orders
but not server-side blocking due to concerns over fraud and abuse.
Thus, server-side blocking and censorship are about
two orthogonal, although not independent, issues:
who is doing the blocking and upon whose orders.

The relationship between our measurements and discrimination is even
more complex.
Our measurements certainly do not distinguish the willful targeting and
harming of a group from implementing needed security precautions, or even from
accidental misconfigurations.
Only in the a subset of cases can we even say that the blocking of a
vantage in a region is because the vantage is in that region.

However, the law also contains notions of indirect discrimination,
such the US's notion of \emph{disparate impact}.
We can draw an analogy between our findings and this notion.
Disparate impact considers large differences between groups receiving
some adverse outcome concerning enough to warrant justification.
Only in the absence of such justification do the differences raise to
the level of being illegal.
Similarly, we believe our findings are concerning from the perspective
of nondiscrimination but not necessarily problematic.
Section~\ref{sec:discussion} considers these issues in more detail.

Needing justifications to put our measurements into context,
we also conducted a more qualitative examination of social media and
web forums to gain some insight into the causes and effects of such
blocking.

Two prominent causes are concerns over security and the costs of
serving traffic to regions with high costs of doing so.
The security concerns include not just hacking but also fraud, such as
Nigerian~419 scams~\cite{419scam}, Amazon delivery
scams~\cite{shippingscam}, and dating scams~\cite{pofscam}.
From the websites' points of view, their services suffer monetary loss
from such abuse, and blacklisting the IP addresses of a region
associated with such abuse provides a low-cost way of reducing such
loss.
We also find that serving traffic to the developing world
can be more expensive than doing so to the developed world, which
could reduce the willingness of websites to serve some regions.

Turning to the effects, we found not only annoyed foreigners\footnote{See, e.g., Vint Cerf on Verizon blocking: \url{https://twitter.com/vgcerf/status/903723705838252032}}, but also 
disrupted online shopping experiences for locals.
More generally, with many key services, such as education, commerce,
and news, offered by a small number of web-based Western companies
that might not view the developing world as worth the costs, these
indiscriminate blanket blocks could slow the growth of blocked
developing regions.

\section{Related Work}

\mct{cite the following works on figuring out where censorship happens: crandall2007conceptdoppler, xu2011internet, verkamp2012inferring, aryan2013internet}

While we believe ours to be the first study dedicated to detecting and
understanding server-side website blocking in the developing world, we
are not the first notice the phenomenon.
Prior work has qualitatively analyzed regional discrimination.
Burrell~\cite{burrell2012invisible} analyzed regional discrimination
from the perspective of two types of actors on the Internet: Internet
users in Ghana (including both ordinary users as well as scammers)
and Western webmasters. 
By observing and interviewing Ghanaian Internet users she identified
confusion that arises from 
the social differences between Ghana and the West.
For example, 
asking for money or gifts from potential partners is socially acceptable in Ghana,
but Ghanaians get blocked on online dating sites for doing so. 
Her analysis of one webmaster forum (\url{WebmastersWorld.com}) showed that Western web admins 
view Africa as a source of Internet abuse without any legitimate users, and often advise other admins to blacklist 
the entire continent for abuse originating from only one region. 

In addition to such qualitative evidence
at least one study measured unavailability from a developing country.
Bischof~et~al.\@ studied Internet connectivity in Cuba, focusing on
routes, performance, and availability~\cite{bischof15imc}.
They found that 111 (2.5\%) of 4,434 tested domains 
(those websites of Alexa top 10k that support HTTPS)
to be unavailable in Cuba.
They find unavailability to be particularly common among websites
related to
 finance,
 ad networks,
 computer hardware, and
 adult content.
Observing that 51 of the 111 (46\%) unavailable domains are also
unavailable in Sudan, they conclude that US sanctions might play a
role given that both countries are subject to them.

Johnson~et~al.\@ study Internet use in the village of Macha,
Zambia~\cite{johnson11www}.
The village network reached the Internet via a proxy and satellite link.
They found that 2.39\% of the proxy responses were HTTP 503 Service
Unavailable, 1.98\% were 504 Gateway Timeouts, and 0.25\% were 502 Bad
Gateway.
They attribute this unavailability to the proxy being overloaded.
Follow-up work found that failure rates remained high after switching
to a faster link with flows of large size being more likely to
fail~\cite{zheleva13dev}.
They attribute their findings to changes in Internet use driving
bandwidth consumption up.

Others have studied issues with Internet connectivity in the
developing world
(e.g.,~\cite{du06www,ihm10nsdr,johnson10nsdr,johnson11www,johnson12ictd,zheleva13dev,chetty13dev,zaki14imc,gupta14pam,kende15internetsociety,fanou15pam,noordally16aintec,fanou16www}).
For example, Zaki~et~al.\@ conducted a measurement study to understand
the high latency of the web in Ghana~\cite{zaki14imc}.
They conclude that the large number of connections needed to load webpages
with redirections and remote content strains the local DNS and caching
infrastructure.
To reach this conclusion, they loaded webpages from both locations
within Ghana and from control locations (e.g., New~York), in a fashion
similar to our study.
Gupta~et~al.\@ come to similar conclusions~\cite{gupta14pam}.

Kende and Rose report that hosting websites in Africa is more
expensive for African websites than hosting them outside of
Africa~\cite{kende15internetsociety}.

Another study of Macha, Zambia,
found that much of the traffic must leave the village over a slow
satellite link to reach services, such as Facebook, despite being a
messages between village residents~\cite{johnson12ictd}.
Fanou~et~al.\@ study content providers in Africa and the routes
between them and users, focusing Google's cache
network~\cite{fanou16www}.
They find poor connectivity between networks within Africa,
sometimes making it faster for an African network to access data
outside of African than that of a different network within Africa.
This finding perhaps explains another of their findings: many websites
local to Africa are hosted outside of Africa.

Given our suspicion that some of the unavailability is from
security-motivated server-side blocking, we find works studying such
blocking of Tor exits relevant.
Khattak~et~al.\@ performed a broad, systematic enumeration and characterization of websites and IP addresses
that treat Tor users differently from normal connections~\cite{khattak16ndss}.
They ran two complementary measurement campaigns:
(1) At the network layer, they scanned the entire IPv4 address space (with a small exclusion list);
(2) At the application layer, they probed the top 1,000 Alexa websites.
At the network layer, they estimated that at least 1.3~million IP addresses that would otherwise allow a TCP handshake
on port~80 blocked the handshake if it originates from a Tor exit node.
At the application layer, on average 3.6\%
of the top 1,000 Alexa websites blocked access from Tor users.

Singh~et~al.\@ further explores this issue, attempting to determine
the motivations and methods behind Tor
blocking~\cite{singh17usenix}.
As with Khattak~et~al.'s study~\cite{khattak16ndss}, they crawl the
web using Tor and without using Tor to detect Tor blocking, but they
also search for additional forms of blocking: websites preventing
logging in or using search functionality.
To understand why blocking happens, they analyze email complaints sent
to exit operators, including for exits they set up, to see what
sorts of abuse webmasters complain about and whether they are
associated with various attributes of the exits.
They also analyzed IP blacklists to understand how Tor exit nodes get
onto them.

Another line of work focuses on measuring specific types of geographic
discrimination on the Internet.
Mikians~et~al.\@ examine price discrimination on e-commerce websites based on
geolocation~\cite{mikians12hotnets,mikians13conext}.
Using Planet~Lab nodes from 6 locations
(New~York, Los~Angeles, Germany, Spain, Korea, and Brazil),
they showed that the price  of certain products, such as e-books, computer games,
and office supplies, differ between buyers at different locations.

Vissers~et~al.\@ looked for, but did not find, price discrimination
for flights based upon location and other attributes (OS, browser, DNT
setting, browsing history, and cookie
settings)~\cite{vissers14hotpets}.

During April 2016, Amazon's Prime Free Same-Day Delivery service was not available in the 
predominantly black neighborhoods in the major cities such as Atlanta,
Boston, and New~York~\cite{ingold16bloomberg}.
Amazon claims that the apparent discrimination is a result of minimizing 
the cost of the same-day service by offering it only to areas with 
with high numbers of Prime users.

Others measured the effects of geolocation on web search
results~\cite{hannak13www,xing14pam,kliman-silver15imc} and
on maps~\cite{soeller16www}.

More generally, researchers have searched for differentiation
based on other attributes.
Hannak~et~al.\ looked price discrimination and
price steering (reordering of items to highlight specific products) on
e-commerce sites based on features including logging in, OS, browser,
and account history~\cite{hannak14imc}.
Their research showed price differences due to logging in (Orbitz and
Cheaptickets),
due to click and purchase history (Priceline),
due to browser (Travelocity and Home Depot),
and due to A/B testing (Expedia and Hotels.com).
Others have looked at ad
personalization~\cite{guha10imc,balebako12w2sp,wills12wpes,sweeney13cacm,liu13hotnets,barford14www,lecuyer14usenix,datta15pets,tschantz15csf,englehardt14man,tramer17eurosp,nath15mobisys,book15arxiv}.

\section{Designing the Experimental Setup}

In this section, we discuss how we design and calibrate our experimental
set-up. In particular, we describe how we eliminate any website unavailability
that can happen due to an artifact of crawler design, as such unavailability is
uninteresting for our study. 

\subsection{Crawler Design}

Our experiments have two design requirements: (i) replicate the browser
behavior experienced by real users and (ii) detailed connection-level error
logging. The former ensures that the blocking we are experiencing is not an
artifact of our crawler, and the latter enables us to differentiate cases of
local network issues and censorship from server-side blocking.

We considered two drivers, Selenium and Python's
Requests package, for
fetching the websites, and compared the two with varying header settings
to enumerate the different reasons that can lead to website
unavailability.

Below we discuss our findings on a sample of Alexa top 500~websites: 
\begin{itemize}
 \item \textbf{Timeout:} We first investigated the effect of the time-out value used.
     We observed a significant reduction in the
     number of websites that time-out (from 43 down to 7), when increasing the time-out value from
     10 to 30 seconds. Increasing the time-out value from 30 to 60s does not offer
     much advantage, but a few websites become accessible at 75s. 

 \item \textbf{User Agent:} We observed at least seven websites being
     unavailable when using the default user-agent for Requests. Six of
     these resulted in a \texttt{403~Forbidden} with the default User Agent,
     but responded with a \texttt{200~OK} when the fetch was attempted
     with Requests using the Firefox User Agent (e.g.,
     \url{theverge.com}, \url{glassdoor.com}, and \url{udemy.com}).
     One website (\url{Redd.it}) responded with a 
     \texttt{429~Too Many Requests}, although we only sent one request.

 \item \textbf{Host header:} We found that whether the Host header contains
     ``www.'' in the domain name has a significant effect on the number of
     websites unavailable. For both Selenium and
     Requests, adding a ``www.'' resulted in a number of websites
     failing at the DNS-level. For some of these this is because the domains already have a
     subdomain prepended to the second-level domain (for example, pages.tmall.com). However, not
     pre-pending ``www.'' does not necessarily work for all the websites. For
     example, target.com and ebc.net.tw return a \texttt{200~OK} when fetched with a ``www.''
     in the header, but return a \texttt{403} when fetched without a ``www.''. To
     investigate this further, we checked the behavior of three major
     browsers and observed inconsistent behavior for when they add ``www.''.
     \mj{commenting this version out: ``and that Chrome would automatically add ``www.'' in cases but Firefox
     and Safari would not.'' This was based on Sadia's finding about Chrome,
 which turned out to be inconsistent with our previous finding. We need to
 investigate the discrepancy, so leaving the wording vague for now.}
     \mct{was:
       ``four major
       browsers (Chrome, Edge, Firefox, and Safari), and found that they send the
       Host header exactly as typed by the user in URL bar.''  So, which is it?  
       What's currently there is what Sadia and I remember.  We don't
       have Edge to even try it.  Maybe this call came from the
       ugrads?}

 \item \textbf{Cookie header (relevant in HTTP redirects):} The default
     behavior in Requests is to not maintain state across HTTP
     requests and to not send a Cookie-Header in 3XX
     redirects.
     Servers could use this behavior to identify our crawler as
     automated and block it.
     Although we did not find evidence of this in our top 500 Alexa
     sample, as a precaution, we perform fetches with a
     \texttt{Session} object, which
     maintains state across requests. 
\end{itemize}

Given these findings, we opted to use Python's Requests package with 
a timeout of 30~seconds and
a User Agent of 
\begin{quote}
\texttt{Mozilla/5.0 (Macintosh; Intel Mac OS X 10\_11\_6) AppleWebKit/537.36
(KHTML, like Gecko) Chrome/66.0.3359.139 Safari/537.36}
\end{quote}
Our crawler attempts to load all pages with HTTP, but follows any automatic
switches to HTTPS.
Given the server-dependent behavior of ``www.'', we first attempt the domain
exactly as it is listed in the list of URLs we used.
If the DNS resolution fails and the URL lacks the ``www.'' prefix,
we try adding it.  If it resolves, we provide the modified URL to the crawler.

\subsection{URL List and Vantage points}
We run the measurements from 16 different vantage points. The machines are
spread across nine countries (Botswana, Bulgaria, Kenya, Pakistan,
South Africa, UK, Ukraine, and USA), and belong to a mix of
network types (institutional, home, VPNs, and cloud). 
Table~\ref{tbl:vantages} lists our vantage points.

\textbf{Website list:} Our list of website URLs 
consists of: (i) the top 500 global
websites, 
(ii) the top 500 regional websites for each of the countries where
we have vantage points as well as for 
Bangladesh (where we attempted to have a vantage), and 
(iii) the top 500 websites according to Alexa in the
following categories: 
news, shopping, business (including banks), 
reference (including educational and universities websites),
science, recreation (including traveling and airline websites),
games, computers (including technology companies websites), and
home (e.g., personal finance). The resulting list contains
7081 unique URLs that form the subject of our study.

\begin{table}
\caption{Vantage Server Locations}
\label{tbl:vantages}
\newcommand{\continent}[1]{}
\begin{tab}{lll}
\continent{No.\@ America}	USA	& Berkeley	& home\\
\continent{No.\@ America}	USA	& Berkeley	& institutional\\
\continent{No.\@ America}	USA	& Berkeley	& institutional VPN\\
\continent{No.\@ America}	USA	& Los Angeles	& VPN\\
\midrule
\continent{Africa}      	Botswana	& Gaborone	& ISP\\
\continent{Africa}      	Kenya   	& Nairobi	& ISP\\
\continent{Africa}      	South Africa	& Johannesburg	& ISP\\
\continent{Africa}      	South Africa	& Johannesburg	& cloud\\
\midrule
\continent{Asia}        	Pakistan	& Lahore	& institutional\\
\continent{Asia}        	Pakistan	& Lahore	& dorm\\
\continent{Asia}        	Pakistan	& Lahore	& home\\
\continent{Asia}        	Pakistan	& Islamabad	& home\\
\midrule
\continent{Europe}      	Ukraine 	& Kiev  	& VPN\\
\continent{Europe}      	Bulgaria	& Sofia 	& VPN\\
\continent{Europe}      	United Kingdom	& London 	& VPN\\
\continent{Asia}        	India   	& Mumbai	& VPN\\
\end{tab}
\end{table}

\section{Finding Regional Blocking}
\label{sec:measurements}

In this section, we discuss our methods for finding regional blocking, and
for the high-confidence candidates, establishing that the blocking happens at
the server-side, as opposed to middlebox interference.

\subsection{Datasets} 

We collected data for each vantage in Table~\ref{tbl:vantages}
three times between
May~18 and 20, 2018, yielding three views of each URL from each vantage (with
the exception of the Islamabad vantage for which we do not have data from the
first day). Our collection logs record the raw HTML response (if present), as
well as meta-information such as HTTP response codes and any errors.

\subsection{Finding Regional Blocking Candidates}

We start with a broad list of websites being unavailable due to any reason, and
then narrow them down to candidates for regional blocking.

\paragraph{Identifying Unavailable Sites}
We identify any site
resulting in an error (such as, TCP~connection error, DNS~error) or serving a
non-20X response as unavailable. 
We also include 200~OK responses that show a blockpage.
To do so, we leverage
block-page identification regexes from our prior work~\cite{khattak16ndss}.

From the raw results, it is easy to find many instances of URLs being
unavailable to a vantage in Table~\ref{tbl:vantages}.
However, we are not interested in mere unavailability, which could be
caused by network failures.
So, we first identify those URLs that are likely blocked
(intentionally unavailable) at a regional scale.
Our later analysis will tease apart different types of blocking and
regional focuses.

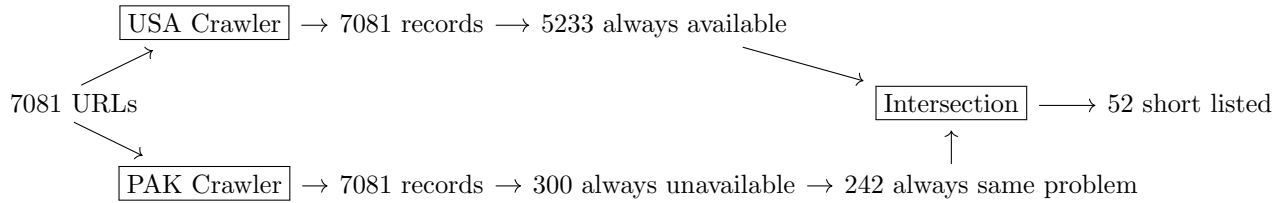
\begin{figure*}
\centering
 \begin{tikzcd}[math mode=false, row sep=1em, column sep=1em] 
&\fbox{USA Crawler}\arrow[r]  & 7081 records \arrow[r] & 5233 always available\arrow[rd]  &       \\ 
7081 URLs\hspace{-6ex}\mbox{}\arrow[ur]\arrow[dr]&&                        &                                  & \fbox{Intersection}\arrow[r] & 52 short listed\\ 
&\fbox{PAK Crawler}\arrow[r]  & 7081 records \arrow[r] & 300 always unavailable \arrow[r] & 242 always same problem\hspace{-7ex}\mbox{} \arrow[u]\\
\end{tikzcd}
\caption{Short-listing URLs.  As an example, we show the results for comparing Pakistan to the US.}
\label{fig:short-listing}
\end{figure*}

\paragraph{Short-listing URLs}
As a first step, we use automated means to short-list a number of URLs
small enough for manual analysis in a way that focuses our attention on
those most likely to be blocked at the regional scale.
Figure~\ref{fig:short-listing} provides an overview of our approach.

To find regional-level differences, and not just network-level ones,
we aggregate results across all the vantages within a region.
We examine three regions: the US,
Pakistan, and Africa.
Given that complaints of regional blocking often come from the developing
world, we particularly focus on finding websites available in the US
but not Pakistan or Africa.

Since our goal is to have high-confidence in our claims that some
websites engage in server-side blocking, of regions in particular, as opposed to finding all instances of
it, we do not treat the two sides of these comparisons equally.
For the US, we aim to underestimate the availability of websites while
for the developing region, we aim to overestimate it.
This difference helps to ensure that we underestimate the number of sites
available in the US but not in the developing region, 
thereby, controlling the false positive rate
for claims of server-side blocking of regions.

With this in mind, to aggregate the results within a region,
we use minimum availability for the US
and the maximum availability for the developing region.
That is, we only short-list URLs that always were available in the US
and never were available in the developing region.
This is a high bar, but in addition to helping to control for false
positives, it captures the intuition that such blocking 
involves blocking whole regions at a time.

Furthermore, to help ensure that the differences were systematic, and
not just the result of network issues, we throw out those URLs that
were not always unavailable in the developing region in the same manner.
For example, we do not short-list a URL that failed to load from 
Botswana due to a DNS failure and from Kenya due to a missing SYN-ACK,
since 
the availability might come from
two different unintentional network issues preventing the connection.
Again, while this is a high bar, it is motivated by the intuition that
region-wide blocking entails a single blanket block of a region, not many
little blocks.
Furthermore, the intent that is suggested from the uniformity of the
unavailability moves the finding from mere unavailability to a
suggestion of active blocking.
However, we acknowledge that, in addition to possibly eliminating from
our short list actual instances of regional server-side blocking that were mixed with
network failures, our short list might include unintentional
unavailability from network failures large enough to affect all
of our vantages in a region.

\paragraph{Manual Checks}
To ensure that our crawler did not introduce artifacts into our data,
we manually loaded each short-listed URL from locations in the
apparently blocked region.
For Pakistan, we loaded them manually with 
the Chrome browser from two vantage points in Lahore, one of them not present
in our crawl measurements.
For Africa, we lacked a physical presence and instead manually
operated Selenium over SSH connections to our Johannesburg cloud vantage point,
which, unfortunately, could introduce artifacts similar to our
automated crawls with the Requests package.

\subsection{Establishing Server-side Blocking}

While the above approach identifies blocking, that is, persistent
(and, thus, probably intentional) unavailability, it cannot
distinguish between server-side
blocking, client-side ISP blocking, and censorship implemented with a middlebox.
In fact, even an explicit block page could be injected by a censor
instead of coming from the requested server.
The short-listed URLs could represent websites censored by each
country in the developing region.
Thus, we take steps to determine
which of the short-listed URLs are unavailable due to server-side
blocking.

Being able to reach the server rules out the quintessential form of
transnational censorship: the client's government blocking of requests
leaving its country.
However, it does leave open two less discussed forms of censorship:
(1) the client country blocking returning responses, and
(2) the server's country blocking connections.
In the second case, the server's government could do so either
directory using its own network infrastructure or indirectly by pressuring
the server into blocking the connection itself.

On the flip side, it is possible that a website implements a
a block by outsourcing the blocking to network infrastructure between
it and the would-be visitors.
In some sense, a website operator using CDN controls is an example of
such, but in such cases, our measurements treat the CDN as the
website's server.
A more tricky case of this is websites purposefully providing
unhelpful DNS records to some DNS resolvers.

With all this in mind, we do not believe it is possible to perfectly
distinguish between server's choices and censorship.
Nevertheless, we further restrict our attention to those cases where
we can determine that server is at least reachable from the vantage.

\paragraph{Traceroute}
To determine whether the server or a middlebox is responding to our
requests, we use traceroute to
determine how far requests for a particular unavailable website make it.
We consider only those URLs that produce consistent traceroute
lengths to have reached the website server. Out approach is inspired by the
work of Xu et al., who use traceroute to determine where in China the censorship occurs~\cite{xu2011internet}.

We first use an ICMP-traceroute to establish a base number of hops that a given
server is away, since with good likelihood the ICMP echo packets do not
experience response-path middlebox interference.
\mct{I don't know what this means or why it is important.}
We compute response-path lengths as
follows: 
If the ICMP traceroute completes, we use the first hope where
responding IP address equals the destination server IP address.
\mct{from DNS? from the response to our Requests request?}\mj{this is from
an altogether new IP from scapy's DNS resolution. Ideally we'd do the
responding IP to our Requests request. For now lets leave the wording as is.}
If the traceroute does not complete with a response from the intended
destination, we underapproximate
the distance to the server by using the length to the last
responding hop. \mj{This underapproximation may cause us to falsely miss
interfering middleboxes (and falsely conclude the blocking is from the true
server) for the cases where the middlebox lies onwards on the path. This is because for
such cases, the
TCP/ HTTP path length will be greater than the ICMP path length. Note if
both TCP and HTTP path length are greater than ICMP path length we always end up concluding that the
blocking is from the server. It is hard to discuss
it here without first discussing our detection heuristics first. }
\mct{say what soundness means}
\mct{at this point what do we know?}

Next, we perform a TCP-SYN traceroute using TCP-SYN packets with
increasing TTLs.
If we receive a response for TCP, 
we use the first hop from where we receive a TCP packet as the length
of the TCP traceroute.
In the case, where we do not receive a response our test ends
inconclusively since on possible explanation is a middlebox
blocking the test but another is the server dropping the packets.
\mct{do we not retry to make sure it's not network issues?\mj{we do, but I am
not entirely sure if we ended up recrunching all the results since the change was added.
I'd have to double-check. There are retries in the
HTTP stateful version too, and different kinds (first for receiving SYN-ACK,
then for receiving a response for the HTTP request packet). I am not sure if it is worth going
going into that much details about retries?}}
\mct{what do we know?}

Given the two traceroutes, we use two methods for identifying middlebox interference:
\begin{enumerate}
\item If the TCP response-path length is shorter than the ICMP path
  length by more than 3~hops, we flag it as likely to be middlebox
  interference.  We use threshold of 3 since the path lengths can
  fluctuate to some extent even in the case of no middlebox
  interference.

\item If the TCP response received is spoofed, we flag it as likely to
  be middle box interference.  We detect a spoofed response when we
  receive an ICMP response and the TCP response for the same hop from
  two different IP addresses.  Note that this check only detects the
  middlebox in cases where it sends an ICMP response from its own IP
  address in addition to sending a spoofed TCP response.
\end{enumerate}

If the TCP traceroute does not find a middlebox using the above methods,
we check using a stateful HTTP traceroute.
We send HTTP GET request packets with increasing TTLs.
Similarly to before, this could end inconclusively if we do not
receive a HTTP response.
We apply the same heuristics used for TCP to compare the HTTP traceroute
to the ICMP traceroute.

If the tests are not inconclusive and we do not find interference from
either the TCP or HTTP test, we
conclude that the server is responding to our requests and the block
is server-side.

Our method can both falsely flag a block as server-side and
miss some server-side blocking.
False positives will arise from middleboxes being 3 or less hops from the
website and not returning an ICMP packet with their true address when they send
a spoofed response (since we fail to detect spoofing in such a circumstance).
While we would like to eliminate these false positives, such close
middleboxes very well may be under the control the website anyhow.
We will fail to confirm server-side blocking when our traceroutes do
not receive the TCP responses needed for the above calculations, such
as from silent server or middlebox blocking.
This approach also does not work websites that implement blocks with
DNS poisoning, although arguably, that is not truly server-side
blocking despite being caused by the website.

\subsection{Whitelisting vs.\ Blacklisting}

Let us distinguish between two types of region-based blocking.
In \emph{whitelisting}, a website aims to serve its content to only
visitors within a targeted region.
In \emph{blacklisting}, a website aims to exclude certain regions.
Strictly speaking, this is again, the sort of property that cannot be
measured since a large enough blacklist will look like a whitelist and
a large enough whitelist will look like a blacklist.
However, we can identify cases where a website appears to only be
reachable within its country, which is a strong suggestion of
whitelisting.
As for identifying likely blacklisting, a website being reachable to
some but not all regions outside of its country with no clear basis
related to the website's purpose suggests blacklisting.
With this in mind, we looked at how websites treat regions other than
two compared in the above methods.

\paragraph{Comparison to Other Regions}
To judge whether the regional blocking we have observed is likely
motivated by whitelisting or blacklisting, we look at measurements
from other regions.

In particular, we use three measurements from across Europe.
We selected Great Britain to be an English-speaking, highly developed
country, unlikely to be blacklisted.
We selected Bulgaria as a country that presents security risks to
websites due to the presence of hackers~\cite{constantin12pcworld}.
We selected Ukraine for similar reasons.
We could see both these countries being blacklisted for security
reasons.
We also included India as a country similar to Pakistan.

Since our vantages in these countries are all VPNs, we must take
findings of unavailability to possibly be indicative not of
geo-blocking, but of VPN blocking.
However, we accept findings of availability as indicating availability.
(We took steps to confirm that the VPNs were actually in the countries
claimed, which was not the case for some VPNs we tried previously.)
For a short-listed URL, we take a URL being available in all the
European locations as a
strong signal of blacklisting and being available only in the US as a
strong signal of whitelisting.
Results in between these extremes are more ambiguous.

\subsection{Limitations}

Despite taking these precautions, false positives remain possible.
At the most fundamental level, the measures we took cannot
distinguish between a web master blocking a region as a region and
blocking the individual IP addresses that make up region after judging
each individually by some other criteria.
However, this does not negate that the server is using server-side blocking
that is affecting a region.
Furthermore, given the vagaries of geo-locating IP addresses, it is
unlikely that even a web master determined to block a whole
country will succeed without overestimating to a great deal.

We only sample the vantages of each
region.
While we took steps to ensure a diversity of vantages, it remains
possible that that rather than whole countries being blocked, just the
networks from which we measured were blocked.
Scaling up our measures to more vantages could address this
limitation.
Nevertheless, we show that some parts of some regions are being
affected by server-side blocking.

On the other hand, by consistently opting to be conservative in our
accusations of server-side blocking, we may have missed many instances of it.
In particular, we will miss instances due to network errors causing
the website to be unavailable in different ways in addition to
server-side blocking.
Similarly, we will also miss server-side blocking using more than one method.
Furthermore, we will miss websites blocking with DNS blocking or
other methods that thwart a connection between researching the website, although it is debatable whether such blocks are truly server-side.

\section{Results}
\label{sec:results}

\input{result-table.tex}

Overall, every vantage point experienced some form of blocking (Table~\ref{tab:all}).
Table~\ref{tbl:results} shows the final results for Pakistan and Africa.

\begin{table}
\caption{Overall Results for Pakistan and Africa.  The number of URLs falling into each category is shown.}
\label{tbl:results}
\centering
\renewcommand{\DTstyle}{\textrm}
\setlength{\DTbaselineskip}{1.5em}
\begin{tabular}{@{}lll@{}}
\toprule
Category & PAK & Africa\\
\midrule
\DTsetlength{0.2em}{0.7em}{0.2em}{0.4pt}{0pt}
\mbox{}\hspace{-1em}%
\begin{minipage}{18em}\dirtree{%
.1 Short-listed URLs.
.2 Manually available.
.3 With \textsc{CAPTCHA}.
.3 With delay.
.3 Without issues.
.2 Not manually available.
.3 Traceroute inconclusive.
.4 DNS error.
.4 No SYN-ACK.
.3 Traceroute conclusive.
.4 Stopped short.
.4 Made it to server.
}\end{minipage}\hspace{-3.5em}\mbox{}%
&
\DTsetlength{0pt}{0pt}{0pt}{0pt}{0pt}
\begin{minipage}{0.4cm}\dirtree{%
.1 \makebox[3ex][r]{52}.
.1 \makebox[3ex][r]{10}.
.1 \makebox[3ex][r]{7}.
.1 \makebox[3ex][r]{3}.
.1 \makebox[3ex][r]{0}.
.1 \makebox[3ex][r]{42}.
.1 \makebox[3ex][r]{28}.
.1 \makebox[3ex][r]{11}.
.1 \makebox[3ex][r]{15}.
.1 \makebox[3ex][r]{14}.
.1 \makebox[3ex][r]{0}.
.1 \makebox[3ex][r]{14}.
}
\end{minipage}
&
\DTsetlength{0pt}{0pt}{0pt}{0pt}{0pt}
\begin{minipage}{0.4cm}\dirtree{%
.1 \makebox[3ex][r]{21}.
.1 \makebox[3ex][r]{2}.
.1 \makebox[3ex][r]{1}.
.1 \makebox[3ex][r]{1}.
.1 \makebox[3ex][r]{0}.
.1 \makebox[3ex][r]{19}.
.1 \makebox[3ex][r]{6}.
.1 \makebox[3ex][r]{0}.
.1 \makebox[3ex][r]{6}.
.1 \makebox[3ex][r]{13}.
.1 \makebox[3ex][r]{0}.
.1 \makebox[3ex][r]{13}.
}
\end{minipage}
\\
\bottomrule
\end{tabular}
\end{table}

\input{us-vs-africa-tr-m.tex}                                                   
\input{pak-vs-usa-tr-m.tex}                                                     
\input{us-vs-pak-tr-m.tex}

As expected, we found evidence that websites in the US engage in
blocking by blacklisting developing regions (Table~\ref{tab:uspak} and Table~\ref{tab:usaf}).
More unexpectedly, by flipping the roles of the US and Pakistan in the
methods described in Section~\ref{sec:measurements},
we found evidence that websites within developing
countries block the US (Table~\ref{tab:pakus}).

Another unexpected pattern was the high degree of blocking of Pakistan
by websites in India.
With only a single VPN vantage in India, we cannot have high
confidence about the unavailability of websites in that country, but we
can confirm with high confidence that Pakistan is blocked by up and
running Indian websites.
This finding could be explained by
the tensions between the countries having lead to tit-for-tat vigilante
hackers from each country attacking websites in the
other~\cite{dmello17india-times}. 
(Relatedly, there are reports of Pakistan censoring Indian
websites~\cite{sholli18express}.)

The results for our black/whitelisting tests were more mixed than we
expected.
For example, \url{peapod.com} works only in the US and in Botswana,
producing a 403 for every other vantage.
We suspect this is a case of whitelisting with a geo-location error
for the ISP-based vantage in Botswana.

The website \url{sunpass.com}, which starts out ``Welcome Florida
Visitors! Planning a vacation to Florida?'' blocks all our vantages
outside the US and Great Britain.  
Perhaps, the website decided that only those from the Anglosphere, narrowly
construed, would like to visit Florida.

\url{hud.gov} could be another such example, with an additional
geo-location error affecting our vantage in Kenya.
This website, about housing from the US government may be an
interesting example of a government deciding to block access as a
server operator, instead of as a censor.
Or, given that all the blocking came from DNS errors, a
misconfiguration.

At the opposite end of the spectrum, 12~websites block only
Pakistan, with three being confirmed server-side blocking.
Nine of them, and all three confirmed server-side blockers, are hosted
in India and may be explained by the aforementioned tensions between
Pakistan and India.
Others may be instances of tight blacklisting due to some previous
issue with Pakistan, such as abuse.
Two websites (\url{pikabu.ru} and \url{williams-sonoma.com}) blocked all and only the vantages in
Africa. Two Pakistani websites (\url{joinpaknavy.gov.pk} and \url{joinpakarmy.gov.pk}) block
all vantage points outside Pakistan. 

We found the following types of blocking:
\begin{enumerate}
\item \textbf{Geo blocking}:
Websites explicitly mention that the country is blocked. 
For example, when loaded from London, \url{jcpenney.com} shows a message saying 
``We are currently unable to
provide a shopping experience for this country.''

\item \textbf{Block page with a way to bypass}:
Websites show a static block page with a \textsc{captcha} or a browser-check page.
In both cases, the websites load after extra work from the client side.

\item \textbf{Block page with a way to bypass}:
Websites show a static block page without any way to bypass. 
These cases include the Akamai block page saying ``Access Denied''.
 
\item \textbf{Non-HTTP errors:}

Some websites block by blocking the DNS request, resetting or aborting connection and 
by responding to the request to cause time out.

\end{enumerate}

\section{Why Blocking Happens}
\label{sec:why}

\begin{figure}[h]                                                              
\centering                                                                      
\includegraphics[width=0.4\textwidth]{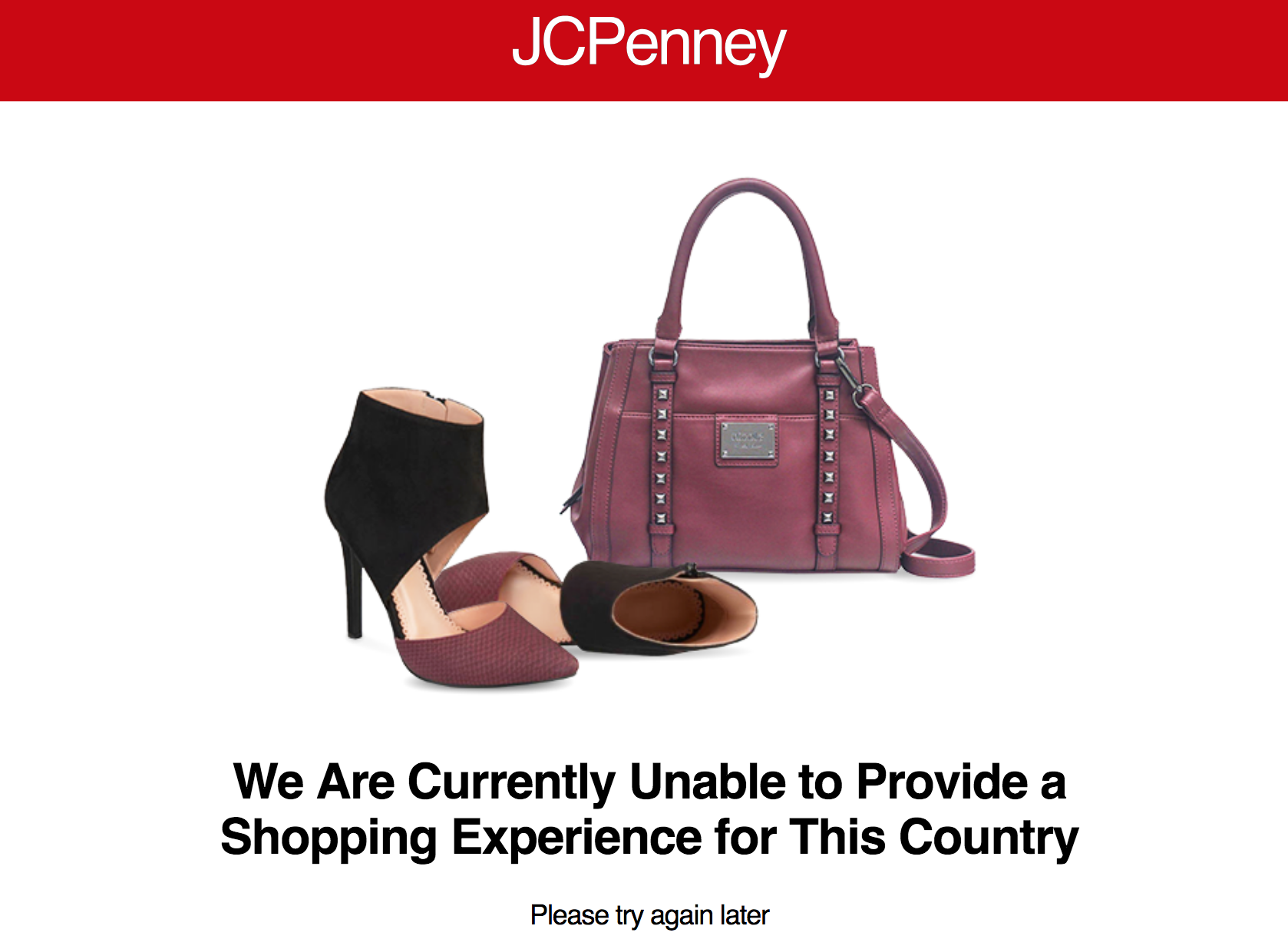}                    
\includegraphics[width=0.4\textwidth]{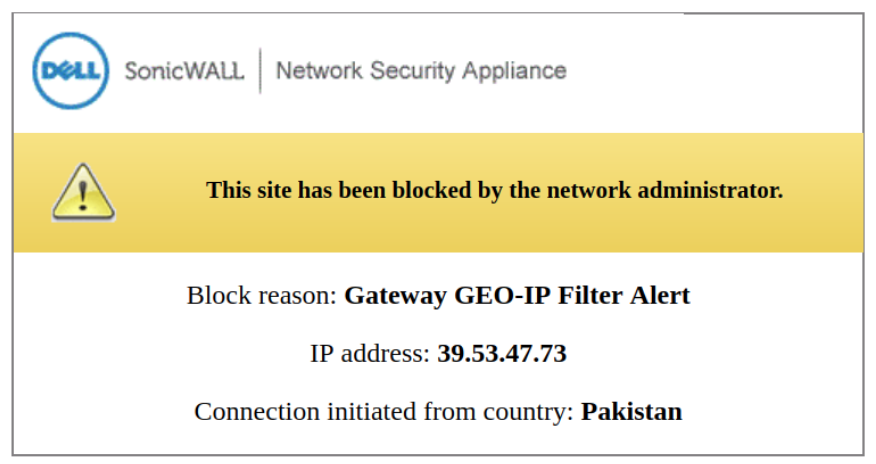}                           
\includegraphics[width=0.4\textwidth]{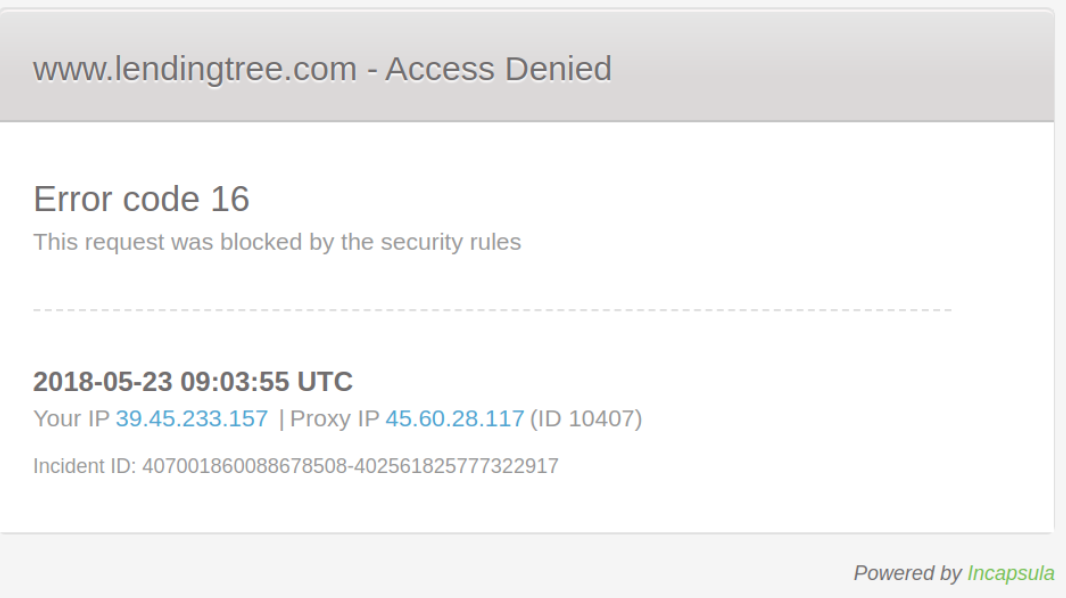}                         
                                                                                
\caption{Example of block pages}                                                                      
\label{fig:blockpages}                                                          
\end{figure}

Numerous possibilities exist for why a website, or content on it, would
be available from Berkeley but not Africa or Pakistan.
Some websites, for example YouTube, explicitly state the reason of
content licence restrictions.
We also found a small number of blockpages that explicitly say that
the reason is securty concerns. For example, \url{lendingtree.com} blocks access 
from Pakistan by security rules (Figure~\ref{fig:blockpages}).
However, often, no reason is provided,
as shown in Tables~\ref{tab:uspak} and Table~\ref{tab:usaf}.
With this in mind, we looked elsewhere for evidence that such blocking
exists.

We first looked for blocking motivated by reducing the costs
associated with security and fraud in particular.
We searched for webpages intended for webmasters that discuss such
blocking.
We found a forum discussion in which the original poster complains of
fraudulent orders using stolen credit cards~\cite{webmaster-world-forum}.
The poster asserts that sometimes the orders are to ``the stolen card
owners real USA address'' and claims that
``They are doing this for no reason other than to be annoying.''
For this reason, the poster says that not shipping to
Africa would not fix the problem and asks
``Is it possible to block african traffic by IP?''
Other forum posters shared technical approaches to blocking web
traffic by country and addition complaints about traffic from some
countries.
For example, one forum poster wrote
\begin{quote}
  In the past, we've had alot of trouble with Nigerian orders.  We've
  had many an order from Lagos, Sweden. But we purchased an annual
  service from StoreIQ to block a list of countries that we
  specify. Haven't had a single Nigerian order since.
\end{quote}
(Lagos is a city in Nigeria, not Sweden.)
Another writes
  ``I nuked most of China about a year ago because of some really
  obnoxious activity on the web site (not necessarily fraudulent
  orders).''
and that 
  ``I can't imagine any reason someone in Nigeria needs to even 
    *see* my web site, let alone place an order on it.''
Others wrote about tricking users into thinking the website is down:
\begin{quote}
  [Our systems] redirect any one
  from a list of chosen countries sent to a page on our site
  which says our site is down. This way they will think something
  is wrong with our site and not come back. It seems to work as
  there is less trafic now from those countires.
\end{quote}
and ``I [\ldots] send all traffic from offending countries
  to a mock 404 page.''
Another poster says
``instead of a 404 we redirect all
  Nigerian traffic the the FBI's Internet Fraud Complaint Center,
  just for fun.''

Another website sets up a discussion of technical methods
of blocking as follows~\cite{gite17nixCraft}:
\begin{quote}
  I admin %
  ecom website and a lot of bogus traffic comes from countries
  that do not offer much in commercial value.  How do I just configure
  Apache or iptables to just refuse connections to certain countries?
\end{quote}
A similar how-to website~\cite{mombrea13itworld} starts 
\begin{quote}
  It's a sad fact that the majority of malicious web traffic to US
  sites originates from the same handful of foreign countries. If your
  site doesn't benefit from actual users living in those countries,
  you may resolve to block them from accessing your site entirely in
  order to prevent repeated attacks.
\end{quote}

A website provides lists of IP addresses by country for blocking
``to minimize on-line fraud, SPAM, floods or brute force
attacks''~\cite{ipdeny18}.

Our search of social media found not only reports of blocking, but
also occasionally a response from the website in question.
President's Choice Financial explicitly cited security as the
motivation~\cite{pcfinancial17twitter}.
Macy's claimed that the reason was not shipping to the country in
question without explaining why that entails
blocking~\cite{macys17twitter}.
(Macy's did not reply when the affected user said she wanted to
ship to a US address.)

From these sources, we conclude that at least some webmasters would
like to block countries due to concerns about fraud or otherwise
undesired traffic originating from them.
Furthermore, it appears common enough that how-to websites provide advice
on doing so.
However, we have not determined what percentage of blocking occurs
for this reason.

We also looked for evidence that websites might be sensitive to the costs
associated with servicing traffic from some countries.
Fastly's content delivery network (CDN) offers points of presence
across the world, including one in South Africa~\cite{fastly18traffic-scalability}.
We found that Fastly charges more for serving requests from some regions
than others.  For example, South Africa is more than twice as
expensive as North~America~\cite{fastly18bandwidth-requests}.
It is possible that websites have decided to not serve traffic to
regions with higher fees and low expectations for revenue.
As with security, we have not attempted to quantify how common this
practice is.

\section{Discussion}
\label{sec:discussion}

Interpreting our results requires care to not jump to unwarranted
conclusions.
We have measured server-side blocking that affects some regions
more than others.
While we have documented some stated causes for such blocking, we have not
quantitatively measured the effect size of each.
Furthermore, except for a small number (at least two, see Figure~\ref{fig:blockpages}) of blockpages
explicitly stating that geo-blocking is the cause, we do not know
which causes lead to the blocking we measured.
Even in these cases, we cannot be sure of the motivations behind the
geo-blocking.
For example, recall the web master posts about using misleading
block pages.
Also, many websites hosted on Akamai or Cloudflare inadvertently 
block Tor users without knowing that they do so
because they use the abuse protection from the content delivery networks (CDNs)~\cite{khattak16ndss}.
Perhaps misconfiguration or poorly understood
configuration led to some of our findings as well.

One may wish to drawn conclusions about issues such as censorship or
discrimination from our findings.
While precisely defining these terms and measuring such high-level
social impacts are beyond the scope of this work, we can
say something about the relationship between our
measurements of server-side blocking of regions and these higher-level
concepts.

For most definitions of \emph{censorship}, our method counts
a strict subset of censorship but also blocking that does meet the definition.
Typically, censorship implies that both parties in a communication are
willing participants and that a third-party (paradigmatically, a
government) disrupts the communication.
Our method counts server-side blocking taken up not only due to orders
issued by a government (such as with economic
sanctions~\cite{bischof15imc}) but also that taken up for voluntary
reasons not involving a third party.
On the other hand, we do not count censorship implemented with
middle-boxes, its quintessential form.

One may wish to say that we have measured some or all of the other
possible voluntary reasons for a server to block a region, such as
asserting that we measured \emph{discrimination}.
However, we only detect blocking for such reasons that is implemented
by server-side blocking.
We would miss any implemented with middleboxes, which, at least in
theory, could happen if the website pays a government to block
requests crossing its boarder headed for the website.

As for discrimination, can we at least say we are measuring the subset
of it implemented as server-side blocking?
The answer depends upon the definition of discrimination used.
For some, such those requiring bigotry or intent, we cannot draw such a
conclusion.

However, we have shown some regions to be disproportionately affected
by a website's use of server-side blocking.
For employment, US anti-discrimination law makes illegal some cases of 
disproportionate impact, namely, those that constitute \emph{disparate impact}.
Disparate impact is a complex legal doctrine.
(For a discussion related to CS, see~\cite{barocas16cal-l-rev}.)
We will not argue that we found disparate
impact, but we do find it helpful to draw an analogy to it.
In particular, we note that an action can lead to disparate impact even when
it and the reasons behind it are facially neutral
towards all groups and do not involve animosity, for example, when it
is correlated with one of the groups.
Furthermore, disparate impact can happen even when not everyone in the
affected group experiences the adverse outcome.
Nor does disparate impact have to be wide spread in the sense of many
employers committing it.
Finally, under the doctrine, disproportionate effects are concerning
enough to warrant a justification and in the absence of such
justification can be problematic (and illegal in that setting).
We also borrow the concept that such justifications should be strong,
in their setting a \emph{business necessity}.

We believe our findings are concerning enough to warrant
justifications in the same sense.
In particular, our study does not determine whether differences in
availability between regions is caused by geo-blocking the affected region
or by some other factor correlated with it.
We also do not show that everyone in the affected region experiences it
nor that it is wide spread across many websites.
However, under our analogy, these issues do not diminish the need for
a strong justification.

As for what would count as a strong justification,
we get little guidance from our analogy since the context of
employment differs greatly from that of serving webpages.
We will leave it to the reader to weigh the justifications
found in Section~\ref{sec:why}.

\section{Conclusion} %

While small in scale, our network measurements provide reasonably
strong evidence that some websites are available from the United
States but not from various locations in Africa and Pakistan
(Section~\ref{sec:results}).
We have also established that at least some websites claim that the
reason for theirs being so is fraud concerns or not shipping to the
location (Section~\ref{sec:why}).
We also found forum discussions relating a desire to block countries
due to concerns over fraud and that traffic can be more expensive in 
Africa (Section~\ref{sec:why}).
However, we cannot attribute all the measured difference in
availability to these reasons.

Despite the uncertainties surrounding the causes of our findings and how to
refer to them, we believe they should be a cause for concern.
The limitations created by such blocks can pose impediments to commerce in developing regions,
either directly by preventing purchases or indirectly by making
examples of successful e-commence sites harder to come by.
 
Furthermore, decisions about blocking remain opaque to the affected users:
for them, things simply do not work, with little explanation as to why.
For example, Akamai-hosted websites show 
a static page to blocked locations simply stating ``Access Denied'',
without any explanation of the underlying issue that led to the blocking,
and with the blog page on the issue being a laundry list of
possibilities~\cite{akamai16blog}.
Users may infer the worst, such as xenophobia, creating a sense of
oppression even if webmasters have legitimate concerns about security.

Even in such cases, wide-net geo-blocking, such as the desire to block
all of Africa we found (Section~\ref{sec:why}), could represent an
overreaction that ignores the great diversity in Africa.
In addition to future research on measuring and understanding the
server-side blocking identified in this work, research could provide webmasters
with tools that allow cost-effective fine-grain blocking in hopes
of preventing such overreactions.

In addition to being interesting in their own right and raising
questions of regional justice, our results have implications for
censorship measurements.
Censorship measurements that do not tease server-side blocking apart from
censorship risks overestimating censorship.
Furthermore, the focus of Western researchers on censorship in the
Eastern world while not acknowledging the server-side blocking existing
in their own countries risks biasing the discussion of openness on the
Internet.

\section*{Acknowledgment}
We thank David Fifield for allowing us to use some of his code.
We thank the ISP that allowed us to use their servers.
We gratefully acknowledge funding support from the National Science Foundation (Grants 1518918 and 1651857) and UC~Berkeley's Center for Long-Term Cybersecurity. The opinions in this paper are those of the authors and do not necessarily reflect the opinions of any funding sponsor or the United States Government.

\bibliographystyle{acm}
\bibliography{censor}
\balance

\end{document}

%% file: preamble.tex
\usepackage{booktabs} %

\newenvironment{tab}[1]
{%
\begin{tabular}{@{}#1@{}}
\toprule
}
{\bottomrule
\end{tabular}}

%% file: result-table.tex
\begin{table*}                                                                  
 \small                                                                         
 \begin{center}                                                                 
\begin{tab}{lllllllllll}
Location/response & 2XX& 403& 4XX (Other)& 503& 5XX (Other)& Conn. Err& DNS Err& Other& Timed Out& Total\\
\midrule
USA, Berkeley, ICSI& 6871.00 &  56.67 &  37.33 &  22.67 &  2.67 &  10.67 &  2.33 &  36.33 &  41.33 &  7081 \\
USA, Berkeley, Home& 6758.00 &  59.67 &  45.67 &  21.67 &  4.67 &  14.00 &  65.67 &  11.00 &  100.67 &  7081 \\
USA, UC Berkeley (VPN)& 6784.67 &  56.67 &  39.33 &  24.00 &  3.00 &  19.67 &  94.00 &  18.00 &  41.67 &  7081 \\
USA, LA (VPN)& 6437.00 &  69.00 &  52.00 &  24.33 &  7.00 &  16.00 &  385.67 &  9.00 &  81.00 &  7081 \\
\midrule

PAK, Lahore (Institutional)& 6218.00 &  96.00 &  43.00 &  28.00 &  2.00 &  4.00 &  3.00 &  129.00 &  558.00 &  7081 \\
PAK, Lahore (Home)& 6595.00 &  113.00 &  56.00 &  28.00 &  3.00 &  16.00 &  152.00 &  40.00 &  78.00 &  7081 \\
PAK, Islamabad (home)& 6689.00 &  109.00 &  53.00 &  26.00 &  3.00 &  50.00 &  54.00 &  10.00 &  87.00 &  7081 \\
PAK, Lahore (Dorm)& 6662.00 &  108.00 &  48.00 &  29.00 &  4.00 &  90.00 &  48.00 &  9.00 &  83.00 &  7081 \\
\midrule                                                                                
ZAF, Johannesburg (ISP)& 6791.00 &  86.67 &  48.00 &  28.67 &  6.67 &  15.33 &  35.00 &  9.33 &  60.33 &  7081 \\
KEN, Nairobi (ISP)& 6802.33 &  83.00 &  49.33 &  24.00 &  5.00 &  17.33 &  35.00 &  9.67 &  55.33 &  7081 \\
BWA, Gaborone (ISP)& 6819.33 &  76.67 &  44.67 &  21.33 &  4.67 &  17.33 &  36.00 &  6.00 &  55.00 &  7081 \\
ZAF, Johannesburg (Cloud)& 6836.00 &  85.33 &  47.33 &  26.33 &  5.00 &  3.33 &  3.67 &  8.67 &  65.33 &  7081 \\
\midrule
BGR, Sofia (VPN)& 6649.33 &  97.00 &  52.67 &  27.67 &  6.67 &  23.33 &  108.00 &  12.00 &  104.33 &  7081 \\
GBR, London (VPN)& 6725.33 &  71.67 &  43.67 &  23.00 &  4.67 &  16.33 &  122.33 &  12.33 &  61.67 &  7081 \\
IND, Mumbai (VPN)& 6467.00 &  85.67 &  45.00 &  25.33 &  4.00 &  25.00 &  44.67 &  8.00 &  376.33 &  7081 \\
UKR, Kiev (VPN)& 6583.33 &  158.00 &  57.33 &  34.00 &  5.00 &  23.67 &  104.67 &  10.67 &  104.33 &  7081 \\

\end{tab}                                                                       
 \end{center}                                                                   
\caption{Crawling result}\label{tab:all}
 \end{table*}

%% file: us-vs-africa-tr-m.tex
\begin{table}
 \small 
 \begin{center} 
\begin{tab}{p{0.86in}p{0.15in}p{0.15in}p{0.15in}p{0.15in}p{0.15in}p{0.15in}p{0.15in}p{0.15in}p{0.15in}p{0.15in}p{0.15in}p{0.15in}} 
URL/Country & BWA& KEN& ZAF& PAK& BGR& UKR& IND& GBR& USA\\ 
 \midrule 
kirklands.com  & \unchecked{403}  & \unchecked{403}  & \trueblock{403}  & \unchecked{403$\dagger$}  & \unchecked{403}  & \unchecked{403}  & \unchecked{403}  & \unchecked{403}  & 200 \\ 
panerabread.com  & \unchecked{403}  & \unchecked{403}  & \trueblock{403}  & \trueblock{403}  & \unchecked{403}  & \unchecked{403}  & \unchecked{403}  & \unchecked{403}  & 200 \\ 
gasbuddy.com  & \unchecked{403}  & \unchecked{403}  & \crawlerblock{403}  & \crawlerblock{403}  & \unchecked{403}  & \unchecked{403}  & \unchecked{403}  & \unchecked{403}  & 200 \\ 
publix.com  & \unchecked{TO}  & \unchecked{TO}  & \unverifiedblock{TO}  & \unverifiedblock{TO}  & \unchecked{TO}  & \unchecked{TO}  & \unchecked{TO}  & \unchecked{TO}  & 200 \\ 
pizzahut.com  & \unchecked{403}  & \unchecked{403}  & \trueblock{403}  & \unchecked{403$\dagger$}  & \unchecked{403}  & \unchecked{403}  & \unchecked{403}  & 200  & 200 \\ 
sunpass.com  & \unchecked{TO}  & \unchecked{TO}  & \unverifiedblock{TO}  & \unverifiedblock{TO}  & \unchecked{TO}  & \unchecked{TO}  & \unchecked{TO}  & 200  & 200 \\ 
www.allconnect.com  & \unchecked{CE}  & \unchecked{CE}  & \trueblock{CE}  & \unverifiedblock{CE}  & \unchecked{CE}  & \unchecked{CE}  & 200  & \unchecked{CE}  & 200 \\ 
forumodua.com  & \unchecked{503}  & \unchecked{503}  & \crawlerblock{503}  & \crawlerblock{503}  & \unchecked{503}  & 200  & \unchecked{503}  & \unchecked{503}  & 200 \\ 
funimation.com  & \unchecked{403}  & \unchecked{403}  & \trueblock{403}  & \trueblock{403}  & \unchecked{403}  & \unchecked{403}  & \unchecked{403}  & 200  & 200 \\ 
tombola.co.uk  & \unchecked{403}  & \unchecked{403}  & \trueblock{403}  & \unverifiedblock{403}  & \unchecked{403}  & \unchecked{403}  & \unchecked{403}  & 200  & 200 \\ 
safeway.com  & \unchecked{403}  & \unchecked{403}  & \trueblock{403}  & \unchecked{403$\dagger$}  & \unchecked{403}  & \unchecked{403}  & 200  & \unchecked{403}  & 200 \\ 
apartments.com  & \unchecked{403}  & \unchecked{403}  & \trueblock{403}  & \unchecked{403$\dagger$}  & 200  & \unchecked{403}  & \unchecked{200$\dagger$}  & 200  & 200 \\ 
duke-energy.com  & \unchecked{DE}  & \unchecked{DE}  & \unverifiedblock{DE$\dagger$}  & \unverifiedblock{DE}  & \unchecked{DE}  & \unchecked{DE}  & 200  & 200  & 200 \\ 
cgg.gov.in  & \unchecked{403}  & \unchecked{403}  & \trueblock{403}  & \unverifiedblock{DE}  & \unchecked{403}  & \unchecked{403}  & 200  & 200  & 200 \\ 
odeon.co.uk  & \unchecked{TO}  & \unchecked{TO}  & \unverifiedblock{TO}  & \unverifiedblock{TO}  & \unchecked{TO}  & \unchecked{TO}  & 200  & 200  & 200 \\ 
lendingtree.com  & \unchecked{403}  & \unchecked{403}  & \trueblock{403}  & \trueblock{403}  & 200  & \unchecked{403}  & 200  & 200  & 200 \\ 
restaurant.com  & \unchecked{CE}  & \unchecked{CE}  & \unverifiedblock{TO$\dagger$}  & \unchecked{CE$\dagger$}  & \unchecked{CE}  & 200  & 200  & 200  & 200 \\ 
pbteen.com  & \unchecked{403}  & \unchecked{403}  & \trueblock{403}  & \unchecked{CE$\dagger$}  & 200  & 200  & 200  & 200  & 200 \\ 
potterybarn.com  & \unchecked{403}  & \unchecked{403}  & \trueblock{403}  & \unchecked{200$\dagger$}  & 200  & 200  & 200  & 200  & 200 \\ 
pikabu.ru  & \unchecked{TO}  & \unchecked{TO}  & \unverifiedblock{TO}  & 200  & 200  & 200  & 200  & 200  & 200 \\ 
williams-sonoma.com  & \unchecked{403}  & \unchecked{403}  & \trueblock{403}  & 200  & 200  & 200  & 200  & 200  & 200 \\ 
\end{tab} 
 \end{center} 
\caption{\small Websites accessible in the US and inaccessible in Africa. \trueblock{red} = Not available manually and true server side blocking, \unverifiedblock{orange} = Not available manually but unverified who is blocking, \crawlerblock{green} = Manually available, \unchecked{yellow} = Not checked manually, *200 = Block page with a 200-OK status code, \textit{200} = The website sometimes loads with a 200-OK status, $\dagger$ = Majority status of the responses, DE = DNS Error, CE = Connection error, TO = time out and RD = Too many redirects.}\label{tab:usaf} 
 \end{table}

%% file: pak-vs-usa-tr-m.tex
\begin{table}
 \small 
 \begin{center} 
\begin{tab}{p{0.8in}p{0.14in}p{0.14in}p{0.14in}p{0.14in}p{0.14in}p{0.14in}p{0.14in}p{0.14in}p{0.14in}p{0.14in}p{0.14in}p{0.14in}} 
 URL/Country & USA& IND& GBR& BGR& BWA& KEN& ZAF& UKR& PAK\\ 
 \midrule 
joinpaknavy.gov.pk  & \unverifiedblock{TO}  & \unchecked{TO}  & \unchecked{CE}  & \unchecked{TO$\dagger$}  & \unchecked{TO}  & \unchecked{TO}  & \unchecked{TO$\dagger$}  & \unchecked{TO}  & 200 \\ 
joinpakarmy.gov.pk  & \unverifiedblock{TO}  & \unchecked{TO}  & \unchecked{CE$\dagger$}  & \unchecked{TO}  & \unchecked{TO}  & \unchecked{TO$\dagger$}  & \unchecked{TO}  & \unchecked{TO}  & 200 \\ 
parfums.ua  & \crawlerblock{403}  & \unchecked{403}  & \unchecked{403}  & \unchecked{403}  & 200  & 200  & 200  & 200  & 200 \\ 
24video.sexy  & \unverifiedblock{403}  & 200  & \unchecked{403}  & 200  & 200  & 200  & 200  & 200  & 200 \\ 
subscene.com  & \crawlerblock{503}  & \unchecked{200$\dagger$}  & 200  & 200  & 200  & 200  & 200  & 200  & 200 \\ 
jrj.com.cn  & \crawlerblock{TO}  & \unchecked{200$\dagger$}  & 200  & 200  & 200  & 200  & 200  & 200  & 200 \\ 
\end{tab} 
 \end{center} 
\caption{\small Websites accessible in PAK and inaccessible in the USA. 
\trueblock{red} = Not available manually and true server side blocking, 
\unverifiedblock{orange} = Not available manually but unverified who is blocking, 
\crawlerblock{green} = Manually available, 
\unchecked{yellow} = Not checked manually, 
*200 = Block page with a 200-OK status code, \textit{200} = The website sometimes loads with a 200-OK status, $\dagger$ = Majority status of the responses, DE = DNS Error, CE = Connection error, TO = time out and RD = Too many redirects.}\label{tab:pakus} 
 \end{table}

%% file: us-vs-pak-tr-m.tex
\begin{table*}
 \small 
 \begin{center} 
\begin{tab}{p{1.2in}p{0.2in}p{0.2in}p{0.2in}p{0.2in}p{0.2in}p{0.2in}p{0.2in}p{0.2in}p{0.2in}p{0.2in}p{0.2in}p{0.2in}} 
 URL/Country & PAK& UKR& BGR& IND& ZAF& KEN& BWA& GBR& USA\\ 
 \midrule 
panerabread.com  & \trueblock{403}  & \unchecked{403}  & \unchecked{403}  & \unchecked{403}  & \trueblock{403}  & \unchecked{403}  & \unchecked{403}  & \unchecked{403}  & 200 \\ 
gasbuddy.com  & \crawlerblock{403}  & \unchecked{403}  & \unchecked{403}  & \unchecked{403}  & \crawlerblock{403}  & \unchecked{403}  & \unchecked{403}  & \unchecked{403}  & 200 \\ 
publix.com  & \unverifiedblock{TO}  & \unchecked{TO}  & \unchecked{TO}  & \unchecked{TO}  & \unverifiedblock{TO}  & \unchecked{TO}  & \unchecked{TO}  & \unchecked{TO}  & 200 \\ 
forumodua.com  & \crawlerblock{503}  & 200  & \unchecked{503}  & \unchecked{503}  & \crawlerblock{503}  & \unchecked{503}  & \unchecked{503}  & \unchecked{503}  & 200 \\ 
funimation.com  & \trueblock{403}  & \unchecked{403}  & \unchecked{403}  & \unchecked{403}  & \trueblock{403}  & \unchecked{403}  & \unchecked{403}  & 200  & 200 \\ 
sunpass.com  & \unverifiedblock{TO}  & \unchecked{TO}  & \unchecked{TO}  & \unchecked{TO}  & \unverifiedblock{TO}  & \unchecked{TO}  & \unchecked{TO}  & 200  & 200 \\ 
www.allconnect.com  & \unverifiedblock{CE}  & \unchecked{CE}  & \unchecked{CE}  & 200  & \trueblock{CE}  & \unchecked{CE}  & \unchecked{CE}  & \unchecked{CE}  & 200 \\ 
peapod.com  & \trueblock{403}  & \unchecked{403}  & \unchecked{403}  & \unchecked{403}  & \unchecked{403}  & \unchecked{403}  & 200  & \unchecked{403}  & 200 \\ 
tombola.co.uk  & \unverifiedblock{403}  & \unchecked{403}  & \unchecked{403}  & \unchecked{403}  & \trueblock{403}  & \unchecked{403}  & \unchecked{403}  & 200  & 200 \\ 
cgg.gov.in  & \unverifiedblock{DE}  & \unchecked{403}  & \unchecked{403}  & 200  & \trueblock{403}  & \unchecked{403}  & \unchecked{403}  & 200  & 200 \\ 
northerntool.com  & \crawlerblock{403}  & \unchecked{403}  & \unchecked{403}  & \unchecked{403}  & \unchecked{403}  & 200  & 200  & \unchecked{403$\dagger$}  & 200 \\ 
odeon.co.uk  & \unverifiedblock{TO}  & \unchecked{TO}  & \unchecked{TO}  & 200  & \unverifiedblock{TO}  & \unchecked{TO}  & \unchecked{TO}  & 200  & 200 \\ 
hud.gov  & \unverifiedblock{DE}  & \unchecked{DE}  & \unchecked{DE}  & \unchecked{DE}  & \unchecked{DE$\dagger$}  & 200  & \unchecked{DE}  & 200  & 200 \\ 
duke-energy.com  & \unverifiedblock{DE}  & \unchecked{DE}  & \unchecked{DE}  & 200  & \unverifiedblock{DE$\dagger$}  & \unchecked{DE}  & \unchecked{DE}  & 200  & 200 \\ 
uhaul.com  & \unverifiedblock{TO}  & \unchecked{TO}  & 200  & \unchecked{TO}  & \unchecked{200$\dagger$}  & 200  & 200  & \unchecked{TO}  & 200 \\ 
lendingtree.com  & \trueblock{403}  & \unchecked{403}  & 200  & 200  & \trueblock{403}  & \unchecked{403}  & \unchecked{403}  & 200  & 200 \\ 
tigerdirect.com  & \trueblock{403}  & \unchecked{403}  & \unchecked{403}  & 200  & 200  & \unchecked{403}  & \unchecked{403}  & 200  & 200 \\ 
payless.com  & \unverifiedblock{DE}  & \unchecked{DE}  & \unchecked{DE}  & 200  & \unchecked{200$\dagger$}  & 200  & \unchecked{DE}  & 200  & 200 \\ 
orvis.com  & \crawlerblock{403}  & \unchecked{403}  & \unchecked{403}  & \unchecked{403}  & 200  & \unchecked{403}  & 200  & 200  & 200 \\ 
www.home-barista.com  & \crawlerblock{503}  & \unchecked{503}  & \unchecked{503}  & 200  & \unchecked{503}  & 200  & 200  & 200  & 200 \\ 
moneytalksnews.com  & \crawlerblock{403}  & 200  & \unchecked{403}  & \unchecked{403}  & \unchecked{403}  & 200  & 200  & 200  & 200 \\ 
ap.gov.in  & \unverifiedblock{DE}  & 200  & \unchecked{TO}  & \unchecked{200$\dagger$}  & 200  & 200  & 200  & 200  & 200 \\ 
telegra.ph  & \unverifiedblock{TO}  & 200  & 200  & \unchecked{TO}  & 200  & 200  & 200  & \unchecked{200$\dagger$}  & 200 \\ 
poloniex.com  & \crawlerblock{403}  & \unchecked{403$\dagger$}  & 200  & \unchecked{403}  & 200  & 200  & 200  & 200  & 200 \\ 
expekt.com  & \trueblock{RD}  & 200  & \unchecked{RD}  & \unchecked{RD$\dagger$}  & 200  & 200  & 200  & 200  & 200 \\ 
www.playstation.com  & \trueblock{404}  & 200  & 200  & 200  & 200  & \unchecked{404}  & \unchecked{404}  & 200  & 200 \\ 
idbibank.co.in  & \unverifiedblock{TO}  & 200  & 200  & \unchecked{200$\dagger$}  & 200  & 200  & 200  & 200  & 200 \\ 
dish.com  & \trueblock{TO}  & \unchecked{TO}  & 200  & 200  & 200  & 200  & 200  & 200  & 200 \\ 
bankbazaar.com  & \trueblock{403}  & \unchecked{403}  & 200  & 200  & 200  & 200  & 200  & 200  & 200 \\ 
brickset.com  & \crawlerblock{403}  & \unchecked{403}  & 200  & 200  & 200  & 200  & 200  & 200  & 200 \\ 
nazk.gov.ua  & \unverifiedblock{DE}  & 200  & 200  & \unchecked{DE}  & 200  & 200  & 200  & 200  & 200 \\ 
tax.virginia.gov  & \unverifiedblock{DE}  & \unchecked{DE}  & 200  & 200  & 200  & 200  & 200  & 200  & 200 \\ 
wipro.com  & \unverifiedblock{TO}  & 200  & \unchecked{200$\dagger$}  & 200  & 200  & 200  & 200  & 200  & 200 \\ 
carandclassic.co.uk  & \trueblock{*200}  & 200  & 200  & \unchecked{*200}  & 200  & 200  & 200  & 200  & 200 \\ 
indianbank.net.in  & \unverifiedblock{CE}  & 200  & 200  & \unchecked{200$\dagger$}  & 200  & 200  & 200  & 200  & 200 \\ 
revenue.alabama.gov  & \unverifiedblock{TO}  & \unchecked{TO}  & 200  & 200  & 200  & 200  & 200  & 200  & 200 \\ 
mahaonline.gov.in  & \trueblock{*200$\dagger$}  & 200  & 200  & 200  & 200  & \unchecked{*200}  & 200  & 200  & 200 \\ 
hotukdeals.com  & \crawlerblock{403}  & 200  & 200  & \unchecked{403}  & 200  & 200  & 200  & 200  & 200 \\ 
iasri.res.in  & \unverifiedblock{DE}  & \unchecked{DE}  & 200  & 200  & 200  & 200  & 200  & 200  & 200 \\ 
lpu.in  & \unverifiedblock{TO}  & 200  & 200  & 200  & 200  & 200  & 200  & 200  & 200 \\ 
anandabazar.com  & \trueblock{403}  & 200  & 200  & 200  & 200  & 200  & 200  & 200  & 200 \\ 
x-minus.me  & \crawlerblock{503}  & 200  & 200  & 200  & 200  & 200  & 200  & 200  & 200 \\ 
uidai.gov.in  & \unverifiedblock{DE}  & 200  & 200  & 200  & 200  & 200  & 200  & 200  & 200 \\ 
snapdeal.com  & \trueblock{403}  & 200  & 200  & 200  & 200  & 200  & 200  & 200  & 200 \\ 
vodafone.in  & \unverifiedblock{TO}  & 200  & 200  & 200  & 200  & 200  & 200  & 200  & 200 \\ 
incometaxindiaefiling.gov.in  & \unverifiedblock{DE}  & 200  & 200  & 200  & 200  & 200  & 200  & 200  & 200 \\ 
ukwhitegoods.co.uk  & \unverifiedblock{CE}  & 200  & 200  & 200  & 200  & 200  & 200  & 200  & 200 \\ 
atariage.com  & \unverifiedblock{TO}  & 200  & 200  & 200  & 200  & 200  & 200  & 200  & 200 \\ 
telangana.gov.in  & \unverifiedblock{TO}  & 200  & 200  & 200  & 200  & 200  & 200  & 200  & 200 \\ 
telegraphindia.com  & \trueblock{403}  & 200  & 200  & 200  & 200  & 200  & 200  & 200  & 200 \\ 
t.me  & \unverifiedblock{CE}  & 200  & 200  & 200  & 200  & 200  & 200  & 200  & 200 \\ 
toto.bg  & \unverifiedblock{TO}  & 200  & 200  & 200  & 200  & 200  & 200  & 200  & 200 \\ 
\end{tab} 
 \end{center} 
\caption{\small Websites accessible in the US and inaccessible in Pakistan. \trueblock{red} = Not available manually and true server side blocking, \unverifiedblock{orange} = Not available manually but unverified who is blocking, \crawlerblock{green} = Manually available, \unchecked{yellow} = Not checked manually, *200 = Block page with a 200-OK status code, \textit{200} = The website sometimes loads with a 200-OK status, $\dagger$ = Majority status of the responses, DE = DNS Error, CE = Connection error, TO = time out and RD = Too many redirects.}\label{tab:uspak} 
 \end{table*}